# The Effect of Increased Access to IVF on Women's Careers

Lingxi Chen[1], Middlebury College

May 2022

## Abstract

Motherhood is the main contributor to gender gaps in the labor market. IVF is a method of assisted reproduction that can delay fertility, which results in decreased motherhood income penalty. In this research, I estimate the effects of expanded access to in vitro fertilization (IVF) arising from state insurance mandates. I use a difference-in-differences model to estimate the effect of increased IVF accessibility for delaying childbirth and decreasing the motherhood income penalty. Using the fertility supplement dataset from the Current Population Survey (CPS), I estimate how outcomes change in states when they implement their mandates compared to how outcomes change in states that are not changing their policies. The results indicate that IVF mandates increase the probability of motherhood by 38 by 3.1 percentage points ($p < 0.01$). However, the results provide no evidence that IVF insurance mandates impact women's earnings.

---

[1] I would like to thank Professor Caitlin Myers, my thesis advisor, for the help and guidance throughout the entire process of this thesis, and the Middlebury College Department of Economics for providing the opportunities and resources necessary to complete this project. I am also thankful for CPS data harmonized and published by Ipums.

## I. Introduction

The gender income gap has been narrowing in the United States and developed countries from the 1970s to the 2000s. However, it has been stagnant in recent two decades, and there is still about a 20% gap between men's and women's weekly and annual earnings (Blau and Kahn 2016). Becoming a parent widens the gender gap in earnings between men and women, that is mothers' income decreases after their first child is born while fathers' income remains similar or increases slightly, and this corresponding income gap can largely be explained by the event of becoming a parent (Kleven et al 2018; Angelov, Johansson, Lindahl 2016). In Denmark, the long-run child penalty in earnings shows that mothers earn 20% less than fathers, which explains most of the remaining gender inequality (Kleven et al. 2018). Analysis of Sweden, Germany, Austria, the United Kingdom, and the United States suggested similar results but different magnitudes (Kleven et al. 2018). A study of gender gaps between couples also demonstrated that 15 years after the first child has been born, the male-female income gap has increased by 32 percentage points (Angelov, Johansson, Lindahl 2016).

The gender gap due to becoming parents that negatively impacts women is known as the motherhood income penalty (Buckles 2008), and it has a greater social impact beyond gender inequality of income for employed women, such as lowering labor force participation rates for women, reducing the prestige of women's careers (Aisenbrey, Evertsson, & Grunow, 2009; Anderson, Binder, & Krause, 2003; Gangl & Ziefle, 2009; Jacobsen & Levin, 1995), and creating a cycle that negatively affects daughters' career outcomes (Kleven et al. 2018). Women with children are less likely to be employed and are more likely to have compromised careers than childless women because they leave the labor force for an extended period or reduce to part-time employment, choose more family-friendly occupations that have lower earning potentials, and decline promotions because of time constraints. Motherhood income penalties have increased in the last forty years and are transmitted through generations, from parents to daughters, suggesting an influence of childhood environment on gender identity (Kleven et al. 2018). Reducing the motherhood income penalty can increase women's participation rate in the labor force, increase women's income, and decrease the overall gender gap.

One way to reduce the motherhood income penalty is to delay mothers' age of having their first child. Studies show that women who become mothers at younger ages and have more children are more likely to accommodate their market work, for example leaving the workforce, reducing hours, and forgoing promotions, so they are more likely to experience greater motherhood income penalties than women who are older at the time of their first births and have fewer children (Blackburn, Bloom, & Neumark, 1993; Chandler, Kamo, & Werbel, 1994; Miller 2011). However, it is not fully established if young motherhood causes low income, or if low income causes young parenthood. Researchers discovered an income premium for women who delay their first birth. Economist using miscarriages to estimate the timing of first birth reveals that an additional year of delay is associated with a 3 percent increase in wage rates and a 10 percent increase in earnings (Miller 2007). Earnings for high-skilled women start to plateau at the point of first birth, which results in considerably higher lifetime earnings for women who delay childbearing (Ellwood, Wilde, and Batchelder 2004). One possible explanation for such a premium is that other observable characteristics, especially education and experience are correlated with age at first birth (Buckles 2008). Buckles' (2008) results support a human capital-based explanation for the delayed premium, but it is inconclusive because most of the observed characteristics, such as the number of children, age at first marriage, occupational choice, and experience, are at least partially endogenous, that they are all affected by age at first birth (Buckles 2008). It is not clear if education causes women to delay or if there is any causal relationship between education attainment and delaying (Buckles 2008). Either possible relationship suggests that delaying first birth may lead to higher income.

The mechanism of how exactly delaying childbearing decreases the motherhood penalty is unclear, but since mothers with older ages at first births tend to have higher education levels and other human capital investments, methods that grant women the choice to delay childbearing can increase women's human capital investment thus decrease motherhood income penalty. In vitro fertilization (IVF), an assisted reproductive technology (ART), achieves this delay by the process of combining a man's sperm and a woman's eggs outside of the body in a laboratory dish, then transferring the embryo (the fertilized egg) into the woman's uterus. The method was first developed in 1978 and has since advanced, achieving about 40 to 50% success rate per cycle for women less than 35 years old (Eskew, Ashley M., and Emily S. Jungheim 2017), which is a relatively low success rate considering the long process. There are millions of IVF births worldwide and IVF births contribute to 1–3% of all births every year in the U.S (Eskew et al., 2017), which shows that despite the low success rate per cycle, women are choosing this procedure. IVF extends the biological timing constraint of declining fecundity and increases the possible age of women at first birth. In the UK in 2018, the average age of first birth for all women who have at least a live birth is 28.9 years old (Statistica), while the average age of an IVF patient was 35.3 years old (Human Fertilisation & Embryology Authority).

However, the high cost of IVF makes it inaccessible in the US. A single IVF cycle—defined as ovarian stimulation, egg retrieval, and embryo transfer—can range from $15,000 to $30,000 (Forbes Health, Conrad 2021), which makes even one cycle unaffordable for most people to pay out of pocket, considering that the median household income is $67,521 in 2020 (US Census Bureau). Moreover, since one cycle's success rate is low, women often need to go through multiple cycles, which increases the total cost of having a child through IVF. Among 330,773 cycles performed in the US in 2019, there were only 77,998 live births, which is about a 24% success rate (CDC's 2019 Fertility Clinic Success Rates Report). The success rate lowers with age, so having IVF available to women when they are young has a bigger effect on their fertility plan. In addition to the high cost, insurance companies in the US do not consider many fertility treatments as "medically necessary", so they are not usually covered by private insurance plans or Medicaid programs (Kaiser Family Foundation). The insurance that covers some fertility treatments mostly only covers fertility testing but not IVF (Kaiser Family Foundation). However, nineteen states have passed fertility insurance coverage laws since 1987, and thirteen of those laws include IVF coverage (National Infertility Association). All thirteen states mandated that IVF must be covered in fully-insured private insurance, but there are variations among each state (National Conference of State Legislatures).

In this research, I use a standard difference-in-differences model to analyze the effect of varying accessibility of IVF in different states on women's fertility and labor force outcomes in the US. For my model, I use the dataset from the Current Population Survey (CPS) fertility supplement (IPUMS 2022) with information about the demographics, work, and education of women in different states. I use the policy exposure of IVF insurance mandates for women aged 45-54 at the time of the survey to estimate three fertility outcomes and three labor force outcomes, such as the age of the women at the time of their first child, percent increase of weekly earnings. For state policy, I match each state with its effective year of IVF legislation according to the National Conference of State Legislatures, and then incorporate the policy information to generate an exposure variable for each woman in the fertility sample, indicating the fraction of her life that IVF was covered by insurance in her state. Based on my results, I found little evidence that the increased accessibility of IVF leads to a delay in motherhood or higher earnings for the overall population. However, the estimated effects are larger for more educated women, suggesting a 2.8 percentage point in the probability of having a child ($p < 0.01$) by age 38. One positive interpretation of this result is that an increase in access to IVF allows more women to have children, increasing fertility without further negative effects on women's labor force participation and earnings. In the future, I will try to use data mining (Durairaj et al. 2013; Zhao et al. 2021) or machine learning (You et al. 2021; Zhao et al. 2021) approaches to furtherly analyze the IVF data.

I organized the rest of the paper as follows. Section 2 will review the main body of literature; Section 3 addresses the datasets used for the study and how I included the states' policies into the data; Section 4 discusses the empirical model; Section 5 reviews the analysis and results; Section 6 concludes and addresses some limitations of my study and suggests further research.

## II. Literature Review

### II. 1 Gender Gap and Fertility

The gender wage gap is the average difference in income for working men and women. Although the gender wage gap is narrowing, there is still a large part that is unexplained. Blau and Kahn (2016) used Panel Study of Income Dynamics (PSID) microdata from 1980 to 2010 and found that the gender wage gap declined considerably during this period. In 2014, the average weekly earnings for women were 82% of that for men, and 79% annually (Blau and Kahn 2016). Gender differences in occupation and industry, and discrimination continued to explain much of the gender gap (Blau and Kahn 2016). Conventional human capital variables explained little about the gender wage gap. Adjusting for only human capital, including education, experience, region, and race, explained 70-80% of the observed gender wage gap. Adjusting for human capital variables and unionization, industry, and occupation, explained 79% of the gender wage gap in 1980 and 91% in the later years. The remaining unexplained 9% included discrimination and or other reasons. If some of the explanatory variables such as experience, occupation, industry, or union status have themselves been directly influenced by discrimination, or indirectly through feedback effects, then an unexplained gap could understate discrimination (Blau and Kahn 2016).

In Blau and Kahn's (2016) review, they summarize the considerable empirical evidence that shows a negative relationship between children and women's wages, commonly known as the motherhood wage penalty. The evidence could be casual, but it could also be due to the selection that women with lower wage offers will have lower costs of having children (Blau et al 2016). The causal argument is that the birth of a child may cause a woman to leave the labor force or change to a more "child-friendly" job, thus decreasing her earnings. From the employer's side, they are also less likely to invest in firm-specific training for women of childbearing age, also contributing to decreased income for mothers. Moreover, motherhood may reduce women's productivity since mothers seek flexibility and have time and travel constraints, which would lead to a decrease in earnings that are not visible in observing women and men's incomes. Finally, mothers also face discrimination in the labor market.

Studying the discrimination in the workplace that could contribute to the gender wage gap, Correll et al. (2007) used laboratory and field experiments and found that there was discrimination against women, but not men, based on parental status. Their results showed that the participants and actual employers viewed mothers less favorably than nonmothers who have the same resume except for parental status, but no differences for fathers and nonfathers.

The gender gap is converging over time, but gender inequality still persists in all countries (Kleven et al. 2019). One of the papers that Blau (2016) included that specifically studies the causal relationship between motherhood and income is Kleven et al. (2018), which illustrated that becoming a parent is the main contributor to gender inequality in the labor market. They estimated the impact of children in the labor market on women compared to men in Denmark, and found a 21% motherhood earnings penalty in the long run. This difference can explain most of the unexplained gender wage gap. Comparing 6 different countries, Kleven et al. (2019) found that Austria, Germany have 51% and 61% child penalty for mothers respectively, 44% and 31% for the UK and the US, and 27% for Sweden. Two possible explanations for the motherhood income penalty include government policies about parental leave and child care, and social norms in the countries. They found that increasing parental leave and public child care have little or no effect on the long-run child penalties for women (Kleven et al. 2019). Social norms are hard to measure and interpret, but Kleven et al. (2018) found the persistent effect of the gender wage gap affected daughters' labor market outcomes but not sons' in Denmark.

## II. 2 IVF and Timing of Fertility

IVF is an important assisted reproductive technology (CDC), which can prolong the fertility period for women. Simoni et al (2017) examined the effect of delaying childbirth with the use of ART on women's careers. Evaluating women's careers and attitudes toward reproductive technologies, they found that women who place high importance on career success place more importance on pregnancy planning, are more likely to delay reproduction, and are more accepting of ART than women with less emphasis on their career. Simoni et al (2017) conducted research on the sample of US women aged 25–45 from 2004 to 2007 from the cross-sectional data from the National Survey of Fertility Barriers (NSFB). The survey suggested that 48.8% of women considered a success in work very important, but women who placed less value on career success were less likely to consider pregnancy planning important (Simoni et al 2017).

The result Simoni et al (2017) confirms Buckles' findings that unobserved characteristics, especially education and experience, explain the premium of delaying motherhood, but those characteristics are all affected by age at first birth (Buckles 2008). There is reverse causality between delayed motherhood and higher human capital investment, and it is not conclusive which one causes the other or if they have a causal relationship that both lead to the outcome of higher earnings. Buckles (2008) examines how much of the delayed premium can be explained by differences in human capital among early and late child bearers. Buckles used data from the National Longitudinal Survey of Youth (NLSY) and found a raw return of approximately 3% per year of delay for hourly wages in 2003, but after controlling for other factors, Buckles found that 90% of the return to delay can be explained by differences in education and experience. The motherhood wage penalty increases with time since birth, and these penalties are greatest for high-skilled women, but are reduced if women delay childbirth (Buckles 2008).

Since women in developed countries are delaying childbearing and encountering more fertility problems in the last 30-40 years, technological advancements have made many more options available to individuals with fertility problems. However, because these technologies are expensive, only 25% of health insurance plans in the United States cover infertility treatment before the insurance mandate. Older, more educated women are using infertility treatments more than other groups after the state mandates were in effect because they have more infertility problems and are more likely to have private insurance (Bitler 2012).

Gershoni et al. (2021) exploited a different IVF policy variation occurring in Israel. They used this policy change as a natural experiment and examined how prolonging women's fertility time could affect women's human capital investment choices. After the policy change that made IVF accessible for all women, more women finished college and graduate school. It granted women better labor market outcomes and lessen the gender gap with men. The authors believed that the gender gap is rooted in biological asymmetries that men have a longer fertility period than women, and extending female fertility is a way to narrow the gender gap (Gershoni et al 2021). It is the case for Israel because the insurance coverage is more comprehensive than the IVF insurance coverage in the US. Another difference is that there is a 2–3 year mandatory military service for Jewish men and women in Israel, and they desire a large family size, which means completing college may interfere with a woman's planned reproductive years (Gershoni et al 2021). The median age for women to enter college was 22.5 in 1994 and a quarter of women entering college are over 25 years old. The average total fertility rate at the time was 2.9 (Gershoni et al 2021), much higher than the fertility rate in the US in 1994 of 2.0 (World Bank 2022). Moreover, the average age at first birth for Israeli-born Jewish women was 26, which is at the time that they would be in college, if they start at 23 years old (Gershoni et al 2021). Their situation makes IVF coverage more impactful in Israel than in the US.

Abramowitz (2017) studied the relationship between the price of ART and women's timing of family including marriage and childbearing, using a natural experiment on the variations of the state fertility insurance mandate. Abramowitz estimated the effect ART state mandates have on first marriage and first birth using data from the 1968–2009 Panel Study of Income Dynamics. The result suggested that the mandates delayed marriage and childbearing for younger women, and increased the likelihood of first marriage and first birth for college-educated women. Mandates were associated with an increased likelihood of marriage at ages 25 and older and motherhood within marriage after at ages 30 and older, but not with the delay at younger ages (Abramowitz 2017).

People today are demanding for extending the definition of infertility treatment such as IVF to include LGBTQ people. Most data only examines the infertility of married heterosexual couples so it understates the rate of infertility. Couples seeking further fertility assistance need to be defined as "infertile" in order to receive the insurance-covered IVF treatment. Non-heterosexual couples have to pay out of pocket for six cycles of intrauterine insemination which is $300-$1,000 per cycle without insurance before they can qualify for infertility diagnosis (The New Yorker, Sussman 2019).

## II. 3 Other Fertility Policies

In addition to IVF-specific literature, two papers that inspired the research design of this paper concern state differences between contraceptive and abortion policies and their effects on women's labor force participation. Bailey (2006) used the variation in state consent laws to estimate the effect of birth control pills on the timing of the first births and women's labor force participation. Bailey (2006) found that if the pill was available to the woman before she was 21, then it reduces the likelihood of her having her first child before 22. The increased legal access to the pill increased the number of women in the labor force and the hours worked (Bailey 2006).

However, Myers (2017) provided new evidence that challenged the powers of the pill raised by Bailey (2006) and showed that abortion policy caused a social change in the 1960s and 1970s. Myers used data from both 1979–95 Current Population Survey (CPS) June Fertility Supplements and the 1980 Integrated Public Use Microdata Series (IPUMS) US Census 1 percent sample to show that access to abortion delayed women's marriages and motherhood but not contraceptives. Myers (2017) assigned three possible levels of access to both the pill and abortion: not available, only available to adult women, and "confidential access" for minors and adults. The policy for the accessibility of the pill had little to no effect on the average probabilities of marrying and giving birth at a young age. The pill was not always available for young women and it had a high failure rate, and demographics could change without any intervention of contraceptives (Myers 2017). However, the policy that enabled legal and accessible abortion caused a 34 percent reduction in young age first births, a 19 percent reduction in young age first marriages, and a 63 percent reduction in "shotgun marriages" prior to age 19 (Myers 2017).

## III. Data
### III. 1 IVF Policy

There are 13 states up to 2022 that have fertility laws mandating IVF coverage in effect. I gathered the information on each state that has an IVF policy from the National Conference of State Legislatures (2022). The states with IVF mandate can also be found on Kaiser Family Foundation (2022), and other fertility treatment centers (The National Infertility Association 2022, American Society for Reproductive Medicine 2022). Two states, Maryland and New Jersey, amended their first IVF mandate stating that it is not required for companies with less than 50 people to have insurance that covers IVF. Four states' amendments (CT, MD, NJ, TX) allowed religious exemptions to the original IVF mandate, that companies may decline to provide IVF-covered insurance for religious reasons. Moreover, some states also imposed limitations on the IVF coverage legislation, such as limiting the number of cycles (HI, IL, NJ) required to be covered, or requiring 5 years of infertility (TX) before insurance has to cover IVF (National Conference of State Legislatures 2022). However, these requirements for all states' IVF coverage do not apply to health plans that are administered and funded directly by employers (self-funded plans) which cover 61% of workers with employer-sponsored health insurance (Kaiser Family Foundation 2022). This would lead to a limited effect of IVF state mandates since 61% of people do not have insurance that is ruled under the state mandates, but rather under the federal rules which do not have IVF coverage (Kaiser Family Foundation 2022). Insurance companies that are required to cover fertility treatments have different plans, and some may only cover a portion of the treatment and have a high payment for the clients making it less accessible.

Figure 1 illustrates all 13 states and the year they adopted IVF coverage. The average year of adoption for IVF-covered states is 2005. The earliest year of IVF mandate is 1987, for Texas and Massachusetts. The most recent states to mandate IVF coverage are Colorado 2022, New Hampshire 2020, and Utah 2020. In the future, there could be more states passing fertility legislation.

Another policy I have controlled for is paid leave policy that has state variations. CA's paid leave policy started in 2004, NJ in 2009, RI in 2014, NY in 2018, and WA in 2020 (Bipartisan Policy Center).

**III. 2 CPS Data**

I use the fertility supplement from the Current Population Survey (CPS) in the US from Integrated Public Use Microdata Series (IPUMS). CPS is a monthly survey conducted by the US. Bureau of Labor Statistics and Census Bureau. I chose the CPS fertility supplement for this research because the survey elicits age at first birth for mothers and identifies which state the woman was in at the time of the survey. I use the mothers' age at first birth to estimate when a woman chooses to have a child, and I investigate if this age is different for women who have IVF coverage available than women who do not. The state information allows me to group women in states that have IVF coverage or in the state without IVF coverage. I assign policy exposure by the current state of residence, which introduces some measurement error due to migration. CPS fertility supplement is part of the CPS collection, but it is conducted irregularly only in June, every five years from 1980 to 1995, and every other year from 2012 to 2020. The fertility supplement surveys men and women of all ages and asks them fertility and marriage questions.

The years that mothers' age at first birth is available are 1980, 1990, 1995, 2012, 2014, 2016, 2018, and 2020. Since I am only observing women's fertility behaviors and labor force outcomes, I limited the sample to only females. Moreover, I choose to only observe women aged 38-54, because they are the cohort that is affected by IVF state policy changes and are old enough that most who will give birth have had their first child, which limited my sample size to 118,058. I chose the age range because in the sample 99% of women who ever had a child, had the child before she was 38 years old, and people older than 54 are less likely to have IVF policy exposure and more likely to retire.

Table 1 shows the summary statistics of outcomes and policy exposure. For outcome variables of the total sample, there are 73% of women in the labor force, and 70% currently employed. For the women currently employed, their average weekly earnings adjusted for inflation is $1013 in 2022 dollars. Considering the policy exposure for IVF coverage, 30% of women in the sample live in the 13 states and had at least an IVF mandate before age 50, and the average year that those women who have ever gotten IVF coverage is 1998. Having the state policy in effect after the woman turns 50 would not affect her fertility behavior or plan because it is past the age of her fertility range, I would consider not women who haven't gotten IVF covered at age 50 affected by the policy. For women who have experienced the policy change before she is 50, the average age that IVF was first covered for her is 30. Women who have IVF available to them when they are in their thirties are less likely to be impacted by this change of

state policy because they could have already made their fertility plans in their twenties, so I included variables indicating if IVF was covered at ages 18, 35, and 40. The percentage of women who had IVF coverage at those ages are 4%, 11%, and 14% respectively. To measure the effect of having IVF continuously, I included IVF exposure, which is the fraction of the women's life that IVF coverage was available for her, and the average is 8%.

Figure 2a and 2b present a bar graph and a boxplot, that describe the difference of percentage women employed and working women's median weekly earnings comparing women who have IVF covered at 35 to women who didn't. Figure 1a depicts that 70% of women in the sample who do not have IVF covered at 35 are working, while 72% of the group with IVF coverage at 35 is working, slightly higher than the group without IVF mandate. Figure 1b compares the weekly earnings of those two groups. The median for no IVF coverage is $820 in 2022 dollars, and $948 for the IVF covered group. The higher edge whisker of the box of the policy-treated group is $3000 also higher compared to $2400 for the control group. Both differences in percentage working and weekly earnings are statistically significant ($p < 0.01$).

Finally, controlling for other demographic variables generates a more accurate relationship between policy and outcome for different groups. The IVF will affect women differently depending on their education and race. Table 2 shows the sample's distribution of education and race. 12% of the women in the sample have less than high school education, and 33% have a high school degree. Women who have some college education, and who have at least a bachelor's degree constitute 28 percent of the sample each. Regarding race and ethnicity, 72% of the sample is White, 11% Black, 11% Hispanic, 5% Asian, and Other constitute 2%.

## IV. Econometric model

### IV. 1 Difference-in-differences Model

I estimated the effects of increased IVF access on women's fertility plans and labor force outcomes, using a simple difference-in-difference framework that exploits spatial and temporal variation in IVF state mandates.

In the simple difference-in-difference estimation, I use an Ordinary Least Squares (OLS) regression to estimate the causal effect of state IVF insurance mandates on two groups of dependent (outcome) variables: women's fertility and labor force outcomes. The first group is fertility outcome, shown in Equation (1).

$$Y_{i,s,t} = \beta_0 + \beta_{IVF\ Exposure_{i,s,t}} + \alpha_{age} + \alpha_{yob} + \alpha_{race} + \alpha_{edu} + \alpha_{paid\ leave} + \alpha_s + \alpha_t + \varepsilon_{i,s,t}$$

(1)

where $\beta_{IVF\ Exposure_{i,s,t}}$, the X variable, means the fraction of IVF coverage experienced by each woman between age 18 to 38. For example, if a woman is surveyed in 2012 in a state with an IVF mandate in effect since 1987, and she is 45 at the time of the survey, so her year of birth is 1967 and she would be 20 years old at the time when she started experiencing IVF coverage by insurance in 1987. IVF Exposure for her would be 18 years (age 20-38) out of the 20 years (age 18-38), which is 0.9. IVF Exposure is measured from 0 to 1, so if the woman didn't experience IVF coverage before age 38, her IVF Exposure would be 0, and 1 if she had IVF coverage in her state from 18 to 38 years old or from 18 up to her age at the time of the survey. For fertility outcomes, I estimated 6 Y variables using the same equation, including "Age at 1st Birth" "Natural Log of Weekly Earnings", "Hours Worked per Week" and indicator variables "Had a Child", "Had a Child before 30", and "In the Labor Force" for each individual in the sample. I included demographic controls for age ($\alpha_{age}$) , year of birth ($\alpha_{yob}$), race ($\alpha_{race}$), education ($\alpha_{edu}$), ($\alpha_{paid\ leave}$) and year of the survey ($\alpha_t$). I also included state fixed effects ($\alpha_s$). Moreover, $\varepsilon_{i,s,t}$ refers to the error term.

## IV. 2 Credibility

The validity of this research design depends on the common trends assumption, which in this case is conditional on the covariates, and I have demographic controls that can account for some of the reasons fertility trends may have differed.

A second potential challenge is that exposure to IVF coverage can be identified only by the current state of residence. This introduces classical measurement error, which causes attenuation bias which is a bias toward finding no effect.

Third, the data over sample minorities, so I included population weights accompanying the data to adjust for the sampling bias, which allowed the final results to be representative of the US population.

## V. Results

Table 3 shows the main regression results. Based on the regression results, there is no evidence of the effect of the IVF mandate on women's labor force outcomes since the effects are very small and lack statistical significance. An increase in IVF mandate exposure on average results in 0.1 percentage points decrease in the probability of being in the labor force, which is a very small amount of increase, and its standard error is 0.009 which lacks statistical significance. There's a 10 minutes increase in weekly time worked, which is negligible and not statistically significant with a high standard error of 0.379. For the change in weekly earnings, the IVF policy's effect leads to a 1.1% decrease with a 0.037 standard error, which is not statistically significant either. There are some effects on fertility outcomes. On average, an increase from not having IVF exposure to having IVF exposure results in 3.1 percentage points of increased probability of having a child ($p < 0.01$), and 1.6 percentage points of increased probability of having a child before age 30 ($p < 0.1$). Mother's age at 1st birth increases by 0.03 years or 10 days, which is small and insignificant. This means that IVF allows women to have children when otherwise they may not be able to have children because they were experiencing infertility. Having IVF coverage may mean that women are more likely to use the method because it is more accessible when covered by insurance.

It is possible that there is no evidence of an effect on the overall population because IVF has heterogeneous effects, that only a particular group is affected while the other groups do not experience an effect. From the state policies, we know that only women with private insurance could have IVF covered by insurance so those women are likely to have higher earnings and higher education levels.

Other potential explanations for the general finding that IVF mandates have little effect on women's careers include that residents of the state may not know about their fertility insurance mandate, so they may have policy exposure but are not aware of them. Since I observe that mothers' ages do not increase, and they are more likely to have children, IVF does not delay motherhood for the overall population and does not lead to higher earnings for mothers for the birth cohort of the 1950s-1970s. It could be because IVF mandates are relatively new, some states as recently as this year and we need to observe them after women have experienced IVF coverage for their childbearing years, 20 years after the start date of IVF mandate, and also wait to observe women who are currently in their childbearing years, younger than 38, who are experience IVF coverage to make their fertility decisions.

In order to examine the heterogeneous effects of IVF policies on women with different educational attainments, I estimated the effects of IVF exposure on fertility and labor force outcomes by education level using the same model. Table 4 shows the effect of IVF exposure on different education attainment, which suggests that the IVF mandate has a stronger effect on women with higher education backgrounds for fertility outcomes, and still small to no effect on labor force outcomes. For women with less than a high school degree, there is no change in the probability of having a child and very little change for other outcomes, which is consistent with my hypothesis that the IVF mandate would affect women with higher levels of education. For women with high school degrees, their weekly hours worked decreased by 1.3 hours ($p < 0.05$). Finally, for women with some college or a bachelor's degree, constituting over 50% of the sample, their result is similar to the overall population results. For the group with some college education, there is a 4 percentage points increase in the probability of having a child ($p < 0.01$), 6 percentage points increase in the probability of having a child before 30 ($p < 0.01$). Similarly, for the group with a bachelor's degree, there is a 4 percentage point increase in the probability of having a child ($p < 0.05$).

The results show that women with a higher human capital investment are more likely to benefit from IVF coverage because they are more likely to have private insurance and children later in life. The result is different from my hypothesis because the population affected by IVF policy is small, so there is no effect on the overall population on their labor force outcomes. Comparing the effect in the US to the effect of IVF coverage in Israel, the average mothers' age at first birth in the two countries are different and Israel's coverage is more universal (Gershoni et al 2021). It is possible that with more universal coverage of IVF, there would be a more significant effect on women in the US.

Moveover, there could be differences in fertility and career effects for different races for all women with a Bachelor's degree or more. I estimated for all women with BA or more of different races. Table 5 shows the effect of IVF mandates on education. For Asian and Black people, there is a 9 percentage point increase in the probability of having a child but it lacks statistical significance. For Hispanic women, there is a 2.7 years increase in age at first birth ($p < 0.05$), which is statistically significant, which could mean that Hispanic people are more likely to use IVF to delay age at 1st birth, and there could be other policies coexisting with the IVF mandate that lead Hispanic women to delay motherhood. For White women, there is a 5 percentage point increase in the probability of having a child ($p < 0.05$), and 0.3 years increase in the mother's age at 1st birth but a 3% decrease in the probability of being in the labor force ($p < 0.1$). Overall, there is no evidence of an effect on labor force outcomes, but an effect on increasing fertility for white college-educated women and delayed motherhood for Hispanic women.

The implication of the results is that IVF mandates do not delay women's age at first birth for most women nor increase their earnings for the overall population. However, the result suggests that an increase in access to IVF allows for increasing fertility without reducing women's labor force participation and earnings. Therefore, the IVF mandate is an effective method for allowing more women to have children without decreasing their earnings.

However, it could be because IVF coverage policies are not accessible enough that there's a very limited increase in access to IVF for the overall population and the state mandates seem to be targeted to cover women who are most likely able to afford the treatments out of pocket, that women who have college degrees and high paying jobs with private insurance in a large company. The literature on delaying motherhood and increased earnings suggests that there is reverse causality in delaying and earnings (Buckles 2008; Simoni 2017), and there is reverse causality in this study as well, that women who have a higher earning potential are more likely to have IVF coverage.

In the appendices, I have included regression estimates for different age ranges, 45-54 instead of 38-54. I have also excluded states with weaker policies, such as limited cycles of IVF and proof of infertility for 5 years HI, IL, NJ, TX. There is no significant difference in the results and none are statistically significant, except for 1.6 hours increase in weekly hours worked ($p < 0.1$) when excluding states with weaker policies. This could mean that as time goes on the initial effect of increased fertility would decrease with age.

## VI. Conclusion

In this study, I estimated the effect of IVF state insurance mandates on women's fertility and labor force outcomes using a difference-in-difference model. Based on our result, I found that IVF exposure leads to an increased probability of having a child for the overall population in the US (3 percentage point increase of probability $p < 0.01$), and especially for women with some college (4 percentage point increase of probability $p < 0.01$) or a college education (4 percentage point increase of probability $p < 0.05$), and White college-educated women (5 percentage point increase of probability $p < 0.01$). The results show that accessible IVF helps women have children, and even leads to mothers being more likely to have children before 30 ($p < 0.1$). Although the ongoing effect of IVF mandates is yet unknown because many states' mandates are too recent, the result shows that an increase in access to IVF allows for increasing fertility.

The results also show there is no evidence of an effect of the IVF mandate on women's labor force outcomes. Maybe not enough women are aware of the policy. Women's earnings decrease or do not change because IVF coverage is allowing more women to have children, and thus focus less on their career and market work. Since IVF allows women to have more children, and more likely to have a child before 30, we would not expect to observe higher earnings due to the delaying motherhood. However, I do not observe a decrease in women's earnings, which means that the IVF mandate is an effective method for allowing more women to have children without reducing women's labor force participation and earnings.

**Figure and Table**

Figure 1
Year Adopted IVF Mandate

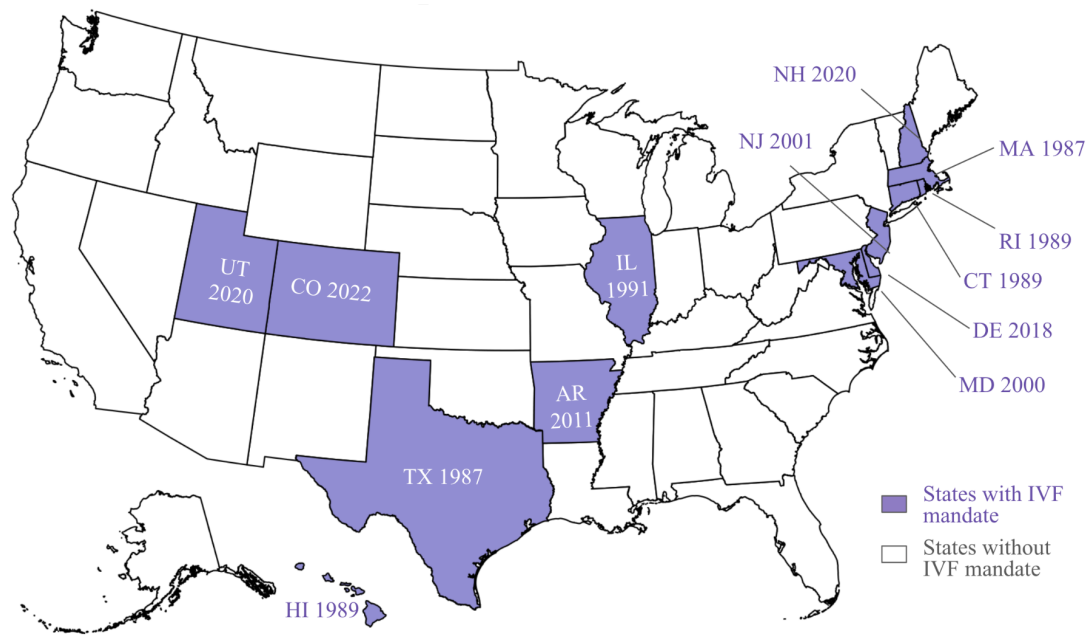

Notes: Policy documented by Chen (2022). Information from National Conference of State Legislatures.

Figure 2a

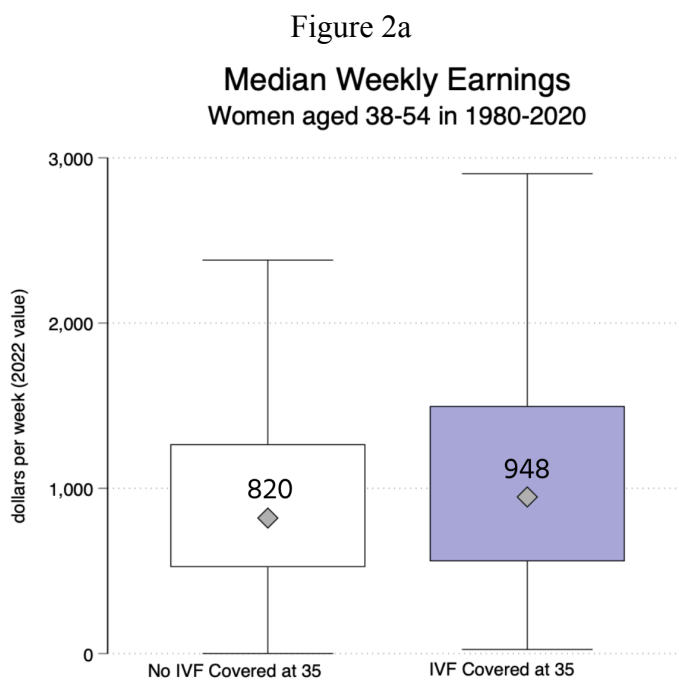

Figure 2b

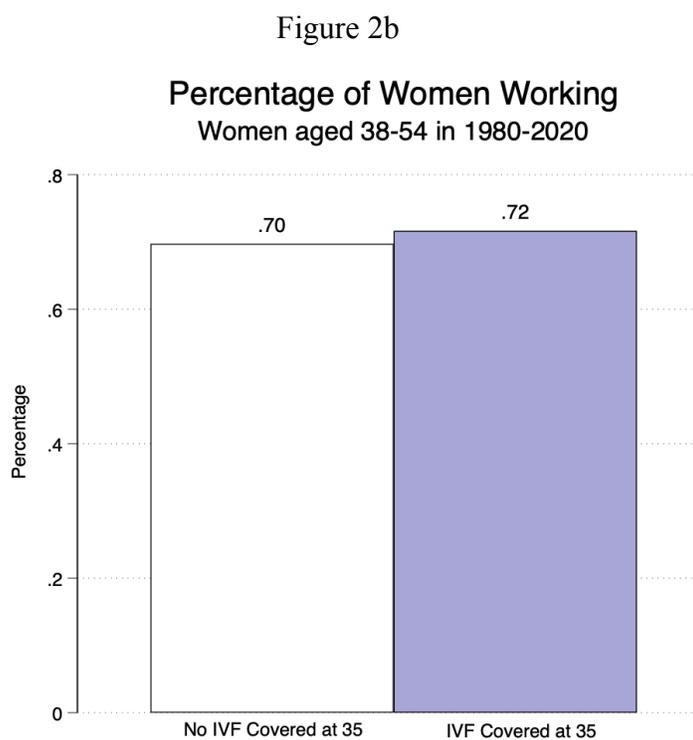

Notes: Bar chart (Figure 1a) describes the percentage of women working by category of if they had or didn't have IVF covered by insurance under state mandate at the age of 35. Box plot (Figure 1b) describes average weekly hours worked for the two groups. Data from CPS fertility supplements.

Table 1
Sample Characteristics: Outcome and Policy Exposure

| Variables | Observations | Mean | SD |
|---|---|---|---|
| **Outcomes** | | | |
| Percentage in the Labor Force | 118,058 | 0.73 | 0.44 |
| Percentage Employed | 118,058 | 0.70 | 0.46 |
| Weekly Working Hours | 57,850 | 38.94 | 10.43 |
| Weekly Earnings in 2022 dollars | 16,616 | 1012.78 | 694.60 |
| **Policy Exposure** | | | |
| Percent have State Coverage | 68,056 | 0.26 | 0.44 |
| Average Year of IVF Coverage | 30,594 | 1998 | 12.80 |
| Age of IVF Coverage before 50 | 21,903 | 30.19 | 12.30 |
| IVF Covered at 18 | 118,058 | 0.04 | 0.20 |
| IVF Covered at 35 | 118,058 | 0.11 | 0.32 |
| IVF Covered at 40 | 118,058 | 0.14 | 0.34 |
| IVF Exposure | 118,058 | 0.08 | 0.25 |

Notes: Labor force outcomes and IVF mandate exposure variations. Data from CPS fertility supplements.

Table 2
Sample Characteristics: Demographics

| Education | Freq. | Percent |
|---|---|---|
| Less than HS | 13,702 | 11.61 |
| High School | 38,813 | 32.88 |
| Some College | 32,882 | 27.85 |
| BA or more | 32,661 | 27.67 |
| Total | 118,058 | 100 |
| **Race** | | |
| White | 85,083 | 72.07 |
| Black | 12,396 | 10.50 |
| Non-white Hispanic | 12,761 | 10.81 |
| Asian | 5,453 | 4.62 |
| Other | 2,365 | 2.00 |
| Total | 118,058 | 100 |

Notes: Education and race categories' frequency and percentage. Data from CPS fertility supplements.

Table 3
Main Regression Estimates: IVF Exposure on Fertility and Labor Force Outcomes
(age 38-54)

| Policy | Fertility Outcome | | | Labor Force Outcome | | |
|---|---|---|---|---|---|---|
| | Ever Had a Child | Mother's Age at 1st Birth | Had a Child Before 30 | In the Labor Force | Weekly Hours Worked | *ln* (Weekly Earnings) |
| IVF Exposure | 0.031*** | 0.033 | 0.016* | -0.001 | 0.175 | -0.011 |
| | (0.007) | (0.119) | (0.009) | (0.009) | (0.379) | (0.037) |
| Number of Observations | 116349 | 85223 | 117978 | 117978 | 57850 | 16616 |

Notes: Linear probability model with year and state fixed effects. Dependent variables are outcomes affected by different levels of IVF exposure. Column 1, 3, 4 are indicator variables, which equals 1 if yes, 0 if no; column 2, 5, 6 are continuous variables. Additional controls include race, education, year of birth, and age.
Robust standard error in parentheses
*** p < 0.01, ** p < 0.05, * p < 0.10

Table 4

Regression Estimates by Education: IVF Exposure on Fertility and Labor Force Outcomes
(age 38-54)

| IVF Exposure By Education | Fertility Outcome | | | Labor Force Outcome | | |
|---|---|---|---|---|---|---|
| | Ever Had a Child | Mother's Age at 1st Birth | Had a Child Before 30 | In the Labor Force | Weekly Hours Worked | *ln* (Weekly Earning) |
| Less than HS | 0.001 | -0.380 | 0.002 | 0.052* | 0.930 | -0.060 |
| | (0.018) | (0.344) | (0.023) | (0.031) | (1.701) | (0.123) |
| | 13290 | 11171 | 13701 | 13701 | 2921 | 1043 |
| High School | 0.013 | 0.021 | 0.025 | 0.009 | -1.303** | -0.036 |
| | (0.012) | (0.215) | (0.016) | (0.018) | (0.646) | (0.071) |
| | 38122 | 29460 | 38804 | 38804 | 15290 | 4790 |
| Some College | 0.042*** | -0.378 | 0.061*** | -0.005 | 0.495 | -0.074 |
| | (0.014) | (0.238) | (0.018) | (0.018) | (0.699) | (0.067) |
| | 32485 | 23647 | 32855 | 32855 | 16747 | 4915 |
| Bachelor Degree or More | 0.042** | 0.391 | -0.013 | -0.025 | 0.831 | 0.037 |
| | (0.017) | (0.270) | (0.021) | (0.017) | (0.665) | (0.071) |
| | 32452 | 20945 | 32618 | 32618 | 22892 | 5868 |

Notes: Linear probability model with year and state fixed effects. Dependent variables are outcomes affected by different levels of IVF exposure listed by race. Column 1, 3, 4 are indicator variables, which equals 1 if yes, 0 if no; column 2, 5, 6 are continuous variables. Additional controls include race, education, year of birth, and age. Robust standard error in parentheses, number of observation for each race under standard error
*** $p < 0.01$, ** $p < 0.05$, * $p < 0.10$

Table 5

Regression Estimates by Race for Women with a Bachelor's Degree or More:
IVF Exposure on Fertility and Labor Force Outcomes (age 38-54)

| IVF Exposure By Race | Fertility Outcome | | | Labor Force Outcome | | |
|---|---|---|---|---|---|---|
| | Ever Had a Child | Mother's Age at 1st Birth | Had a Child Before 30 | In the Labor Force | Weekly Hours Worked | *ln* (weekly earning) |
| Asian | 0.009 | -0.491 | 0.062 | 0.035 | -0.122 | 0.488 |
| | (0.054) | (1.021) | (0.076) | (0.076) | (2.448) | (0.328) |
| | 2606 | 1758 | 2606 | 2606 | 1789 | 471 |
| Black | 0.090 | 0.012 | 0.108 | -0.057 | 0.609 | 0.240 |
| | (0.061) | (1.285) | (0.074) | (0.054) | (2.054) | (0.167) |
| | 2712 | 1757 | 2723 | 2723 | 2032 | 577 |
| Hispanic | 0.018 | 2.728** | -0.097 | -0.090 | 3.042 | 0.313 |
| | (0.068) | (1.351) | (0.090) | (0.076) | (2.453) | (0.407) |
| | 1989 | 1342 | 1997 | 1997 | 1413 | 361 |
| White | 0.049** | 0.341 | -0.028 | -0.033* | 0.827 | -0.034 |
| | (0.019) | (0.283) | (0.024) | (0.019) | (0.792) | (0.083) |
| | 24651 | 15762 | 24793 | 24793 | 17322 | 4363 |

Notes: Linear probability model with year and state fixed effects. Dependent variables are outcomes affected by different levels of IVF exposure listed by race. Column 1, 3, 4 are indicator variables, which equals 1 if yes, 0 if no; column 2, 5, 6 are continuous variables. Additional controls include race, education, year of birth, and age.
Robust standard error in parentheses, number of observation for each race under standard error
*** $p < 0.01$, ** $p < 0.05$, * $p < 0.10$

# Appendices

Table 1

State Insurance Mandates

| State | Year |
|-------|------|
| AR | 2011 |
| CT | 1989 |
| DE | 2018 |
| HI | 1989 |
| IL | 1991 |
| MD | 2000 |
| MA | 1987 |
| NJ | 2001 |
| RI | 1989 |
| TX | 1987 |
| CO | 2022 |
| NH | 2020 |
| UT | 2020 |

Notes: Policy documented by Chen (2022). Information from National Conference of State Legislatures.

Table 2
Regression Estimates: IVF Exposure on Fertility and Labor Force Outcomes
(age 45-54)

| Policy | Fertility Outcome | | | Labor Force Outcome | | |
|---|---|---|---|---|---|---|
| | Ever Had a Child | Mother's Age at 1st Birth | Had a Child Before 30 | In the Labor Force | Weekly Hours Worked | *ln* (Weekly Earnings) |
| IVF Exposure | 0.014 | 0.060 | 0.017 | 0.002 | -0.061 | 0.018 |
| | (0.010) | (0.180) | (0.0013) | (0.014) | (0.525) | (0.056) |
| Number of Observations | 67078 | 43010 | 68033 | 68033 | 33822 | 9390 |

Notes: Linear probability model with year and state fixed effects. Dependent variables are outcomes affected by different levels of IVF exposure. Column 1, 3, 4 are indicator variables, which equals 1 if yes, 0 if no; column 2, 5, 6 are continuous variables. Additional controls include race, education, year of birth, and age.
Robust standard error in parentheses
*** $p < 0.01$, ** $p < 0.05$, * $p < 0.10$

Table 3

Regression Estimates: IVF Exposure on Fertility and Labor Force Outcomes (age 38-54)

Excluding States with Weak Policies  (HI, IL, NJ, TX)

| Policy | Fertility Outcome | | | Labor Force Outcome | | |
|---|---|---|---|---|---|---|
| | Ever Had a Child | Mother's Age at 1st Birth | Had a Child Before 30 | In the Labor Force | Weekly Hours Worked | *ln* (Weekly Earnings) |
| IVF Exposure | 0.018 | 0.368 | 0.010 | -0.005 | 1.613* | 0.132 |
| | (0.016) | (0.290) | (0.021) | (0.020) | (0.832) | (0.085) |
| Number of Observations | 58733 | 37519 | 59566 | 59566 | 29606 | 8208 |

Notes: Linear probability model with year and state fixed effects. Dependent variables are outcomes affected by different levels of IVF exposure. Column 1, 3, 4 are indicator variables, which equals 1 if yes, 0 if no; column 2, 5, 6 are continuous variables. Additional controls include race, education, year of birth, and age.
Robust standard error in parentheses
*** p < 0.01, ** p < 0.05, * p < 0.10